\documentclass[12pt]{iopart}
\usepackage{graphicx}
\usepackage{iopams}
 \begin{document}
\title{ Correlation of high energy photons and charged 
particles in p+p and d+Au collisions at $\sqrt{s}=200$ GeV }
 \author{Subhasis Chattopadhyay (for STAR Collaboration) \\
\it Variable Energy Cyclotron centre, 1/AF Salt Lake, Kolkata 700064} 
\vskip 0.5cm
\begin{abstract}
 Azimuthal correlations between high energy 
photons from $\pi^0$ decay ($p_T^{\gamma} < 10.5$ GeV) and charged hadrons 
($p_T^{hadron}$ above a threshold of 2, 3 or  4 GeV/c) have been measured in p+p and d+Au collisions 
at $\sqrt{s}=200$ GeV. Clear jet-like correlations are observed at 
$\Delta\phi=0$ and 180 degrees, with peak widths decreasing with 
increasing $p_T^{hadron}$. Jet $\langle j_T \rangle $ and $\sqrt{\langle k_T^2 \rangle}$  in d+Au collisions are deduced from the distributions.
 {\it These results are presented in Quark Matter 2004 poster session.}

\end{abstract}
\vskip 0.2cm

Particles produced by fragmentation of jets are seen to be correlated in azimuth and pseudorapidity.
In high energy heavy ion collisions, a measurement of dihadron correlation
functions is the closest approach to obtain the properties of jets.
At RHIC energy, correlation functions have been measured between leading charged particles
 up to $p_T$ of 6 GeV/c and the associated charged particles with $p_T > $ 2 GeV
for pp, dAu and AA collisions [1].

As the trigger particle energy is increased, the strength of the jet-like 
correlations increases and the width of the correlations 
gets narrower.  Increasing the associated particle threshold reduces the 
background in the correlation functions dramatically.  Thus, higher energy 
trigger particles allow jet properties to be inferred from two-particle 
correlations with much less ambiguity.

In this work, energy deposition by photons 
 in STAR Electromagnetic Calorimeters (EMC) are used for tagging
events with high energy photons, thereby enabling us to obtain correlation
functions with photon as trigger particles up to very high energy.
In the present study we measure the azimuthal correlation function between
 triggered photons and the associated charged particles for pp and dAu 
collisions. We study the variation of widths of the correlation 
functions for various energies of triggered particles and for change in $p_T$ of associated particles. 
We also derive mean jet fragmentation transverse momentum
 $\langle j_T \rangle$ and  
{\it r.m.s} parton transverse momentum $\sqrt{\langle k_T^2 \rangle} $ from the 
widths of the correlation functions.

In this analysis we used triggered events from 2003 run 
for pp and dAu collisions.
 In this run EMC covered the region of 0 $ < \eta <1$ with full azimuthal
 coverage.
 Charged tracks with $|\eta| < 1$ are used as associated particles 
for obtaining correlation functions. We used standard cuts used in STAR for selecting high transverse momentum tracks.

 For every event we selected towers with highest energy deposition above selected
 threshold (High Tower), these towers are considered as trigger tower.
 We then calculated $\Delta\phi = |\phi_{trig} - \phi_{associated}|$
 for all associated particles above various thresholds with
 $p_T^{associated} < E_T^{HighTower}$.
 Correlation functions are obtained by calculating,
 $\frac{1}{\epsilon N_{trig}} \frac{dN}{d\Delta\phi}$, where $N_{trig}$ is the number of trigger towers, $\epsilon$ is the 
efficiency of the associated particles at the selected $p_T$ range.

Fig.1 and 2 show the pedestal subtracted correlation functions for 
dAu collisions. Fig. 1 shows the case of two sets of associated particles 
$p_T$ thresholds, (2 GeV/c and 3 GeV/c) for 
 4.5 GeV $< E^{trig} < $ 6.5 GeV.
Fig.2 shows the case for
 two sets of trigger
 particle energy, 4.5 GeV $< E^{trig} < $ 6.5 GeV and
6.5 GeV $< E^{trig} < $ 8.5 GeV.
Pedestal values of the correlation functions
are obtained by taking the averages in the region of $\Delta\phi$ = 1 to 2.
\begin{figure}
\centering
\includegraphics[height=0.35\textheight]{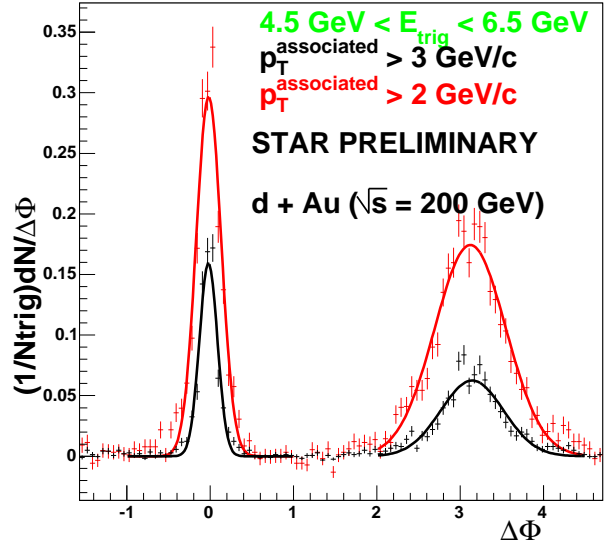}
\caption{ Azimuthal correlation functions for dAu collisions with 4.5 GeV $< E^{trig} < $ 6.5 GeV
for two sets of lower limits in $p_T^{associated}$, 2GeV/c and 3 GeV/c.}
\end{figure}
\begin{figure}
\centering
\includegraphics[height=0.35\textheight]{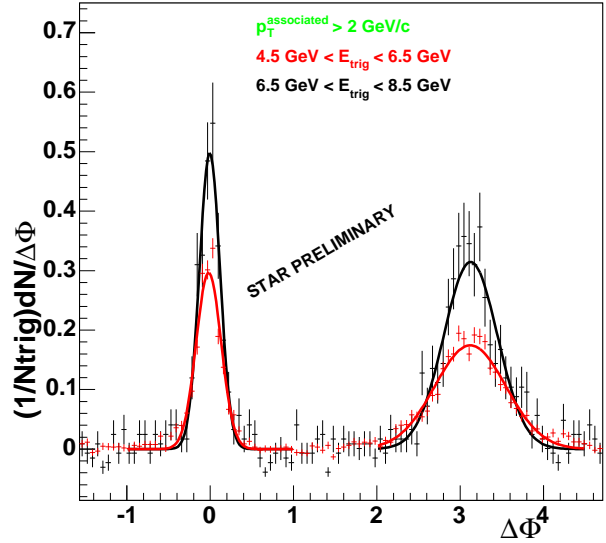}
\caption{ Azimuthal correlation functions for dAu collisions 
for two sets of $ E^{trig}$, 4.5 GeV $< E^{trig} < $ 6.5 GeV and 6.5 GeV $< E^{trig} < $ 8.5 GeV with same lower limit on $p_T^{associated} = 2$ GeV/c.  }
\label{mipcell}
\end{figure}

Two clear peaks are seen. Peaks (near angle peak and far angle peak) 
are fitted with gaussian functions. 
We have studied (a) area under
the peak giving the total number of pairs per trigger tower and (b) $\sigma$ of 
the fitted gaussians on two peaks.
Fig. 3 and fig. 4 shows the variation 
of above mentioned parameters for varying $p_T^{associated}$ and varying $E^{trig}$.
\begin{figure}
\centering
\includegraphics[height=0.4\textheight]{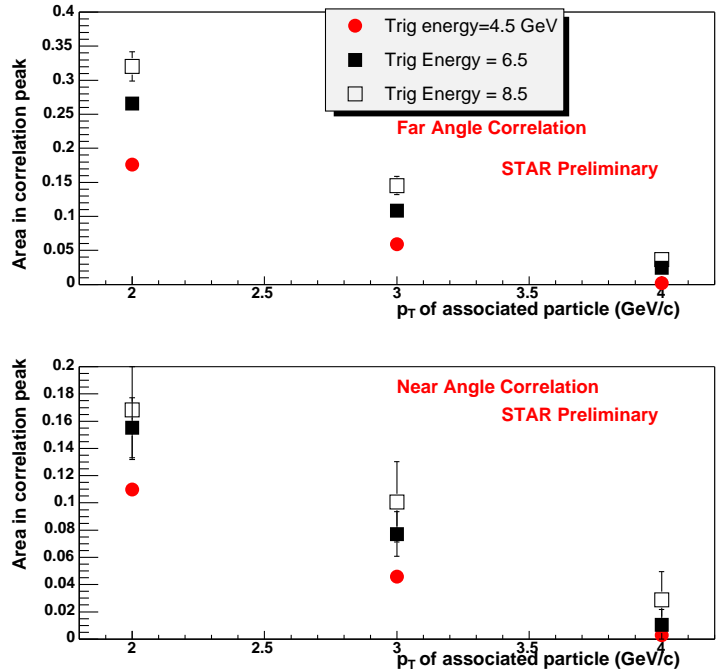}
\caption{ Variation of the area under peaks for near angle ( $\Delta\phi = 0^0$ ) and
 far angle correlation peaks ($\Delta\phi = 180^0$) for various $p_{T}^{associated}$. 
 Three sets of $E^{trig}$ bins are used with 2 GeV width and lower limits of 
4.5 GeV/c, 6.5 GeV/c and 8.5 GeV/c.}
\label{mipcell}
\end{figure}
\begin{figure}
\centering
\includegraphics[height=0.4\textheight]{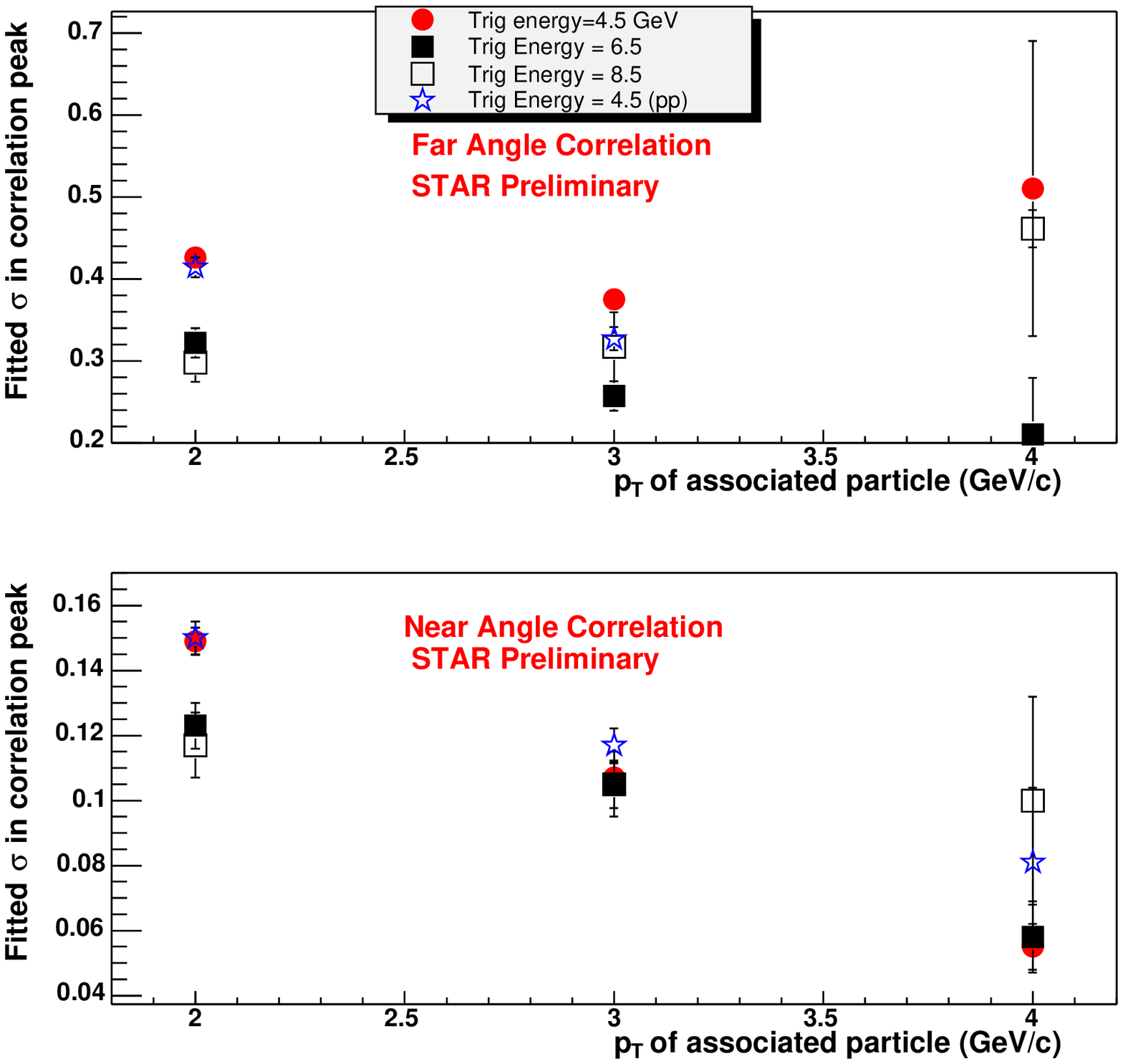}
\caption{ Variation of the fitted $\sigma$ for  near angle ( $\Delta\phi = 0^0$ ) and
 far angle correlation peaks ($\Delta\phi = 180^0$) in dAu and pp collisions for various $p_{T}^{associated}$. 
pp results are shown for 4.5 GeV $< E^{trig} < $ 6.5 GeV.}
\label{mipcell}
\end{figure}

Correlation peaks get narrower with increase in $E^{trig}$ and $p_{T}^{associated}$ as expected from jets.
 Within error pp and dAu shows 
similar widths for 4.5 GeV$ < E^{trig} <$ 6.5 GeV.
 Number of associated particles are larger
for higher $E^{trig}$ and reduces with $p_T^{associated}$. This number 
gives idea about the number of charge particles associated to the
 jets in near and far region.

In order to extend the analysis for jet characterization we calculated $\langle j_T \rangle$ and $\sqrt{\langle k_T^2 \rangle}$ using the formulae,

\begin{eqnarray}
 \sigma_N^2 \approx  \frac{\langle E_{T}^2 \rangle + \langle p_{T}^2 \rangle}{2 \langle
 E_{T}^2 \rangle \langle p_{T}^2 \rangle} \langle j_T^2 \rangle .
 \end{eqnarray}
\begin{equation}
 \langle j_T \rangle = \frac{\sqrt{\pi}}{2} \sqrt{\langle j_T^2 \rangle} .
 \end{equation}
and

\begin{equation}
 \sqrt{\langle k_T^2 \rangle} \approx \frac{\langle E_{T} \rangle}{\langle z
 \rangle} \sqrt{\sigma_F^2 - \sigma_N^2} .
 \end{equation}
where $\langle E_T \rangle$ represents average transverse momentum of the 
triggered towers and $\langle p_T \rangle$ gives average transverse momentum
of the associated particles. $\sigma_F$ and $\sigma_N$ are the fitted widths
of near and far angle peaks.

 The values obtained for $\langle j_T \rangle$ and $\langle z \rangle \sqrt{\langle k_T^2 \rangle}$
are 450 $\pm$ 40 $\pm$ 150 MeV/c and 1.9 $\pm$ 0.2 $\pm$ 0.3 GeV/c respectively
,where $\langle z \rangle $ is the fragmentation function for the trigger photon, and is in the range 0.6-0.8.
 
{\bf References:}

1. C. Adler et.al.,(STAR Collaboration), Phys. Rev. Lett. 90 (2003)082302
\vskip 0.2cm

\end{document}